\documentclass[12pt, hidelinks]{article}
\usepackage[english]{babel}
\usepackage{amssymb,amsmath,amsthm,amsfonts,mathtools}
\usepackage[top=2.8cm, bottom=2.8cm, left=2.8cm, right=2.8cm]{geometry}
\usepackage{graphicx}
\usepackage{hyperref}
\usepackage{natbib}
\usepackage{tikz}
\usepackage{tikz-cd}
\usetikzlibrary{trees}

\newtheorem{theorem}{Theorem}

\newtheorem{lemma}{Lemma}

\newtheorem{proposition}{Proposition}

\newtheorem{definition}{Definition}

\newtheorem{assumption}{Assumption}

\begin{document}
	\title{Exploring the Constraints on Artificial General Intelligence: A Game-Theoretic No-Go Theorem\footnote{I would like to acknowledge the use of GPT-3 models in improving the exposition of this paper, including the creation of the images presented in Figure~\ref{fig:teaser}. This paper was previously entitled ``Political economy of superhuman AI.''}}
\author{Mehmet S. Ismail\footnote{ {\footnotesize Department of Political Economy, King's College London, London, UK. E-mail: mehmet.ismail@kcl.ac.uk}}}
%
\date{This version: 9th November, 2023\\ \href{https://arxiv.org/abs/2209.12346}{First version}: 21st September, 2022}

	\maketitle
\begin{abstract}
The emergence of increasingly sophisticated artificial intelligence (AI) systems have sparked intense debate among researchers, policymakers, and the public due to their potential to surpass human intelligence and capabilities in all domains. In this paper, I propose a game-theoretic framework that captures the strategic interactions between a human agent and a potential superhuman machine agent. I identify four key assumptions: Strategic Unpredictability, Access to Machine's Strategy, Rationality, and Superhuman Machine. The main result of this paper is an impossibility theorem: these four assumptions are inconsistent when taken together, but relaxing any one of them results in a consistent set of assumptions. Two straightforward policy recommendations follow: first, policymakers should control access to specific human data to maintain Strategic Unpredictability; and second, they should grant select AI researchers access to superhuman machine research to ensure Access to Machine's Strategy holds. My analysis contributes to a better understanding of the context that can shape the theoretical development of superhuman AI.
	\end{abstract}

	
\section{Introduction}
\label{sec:intro}

Artificial intelligence (AI) is transforming various domains of human activity, such as healthcare, education, and entertainment. However, as AI systems become more capable and autonomous, they also pose new ethical and societal challenges that require careful consideration and regulation \citep{hadfieldl2017,floridi2019,walz2019,schiff2020,acemoglu2021,cohen2022,london2023}. One of the most pressing and controversial issues is the possibility of creating superhuman AI or artificial general intelligence that can surpass human intelligence and abilities in all domains.

The prospect of superhuman AI has sparked intense debate among AI researchers and practitioners, as well as philosophers, ethicists, policymakers, and the general public. Some view superhuman AI as a desirable and inevitable goal that could bring unprecedented benefits to humanity \citep{bostrom2014}. Others warn of the existential risks and moral dilemmas that superhuman AI could entail \citep{yudkowsky2008,bostrom2014,russell2019}. In 2015, an open letter signed by over 150 prominent AI experts called for more (social science) research on how to maximize the societal benefits of AI systems and ensure the alignment of superhuman AI with human values and interests \citep{horvitz2014}. However, there is still no consensus on whether superhuman AI is feasible or desirable, or how to achieve it safely and ethically \citep{everitt2018}.

In this paper, I adopt a game theoretical perspective to analyze the emergence of superhuman AI, taking into account the social and institutional factors that may influence its development. I propose a framework that captures the strategic interactions between a representative human agent ($H$) and a potential superhuman machine agent ($M$). I consider four idealized assumptions in this framework. The first assumption is Strategic Unpredictability, which captures the game theoretical assumption that $H$'s strategy cannot be perfectly predicted by $M$. The second assumption is Access to Machine's Strategy, which allows $H$ to access the strategy of $M$ in a way that is similar to using a chess engine while playing chess. The third assumption is Rationality, which means that the human agent chooses the strategy that maximizes their payoff, given the strategy of $M$. The fourth assumption is Superhuman Machine, which implies that $M$ can outperform $H$ in every potentially non-zero-sum two-person contest played between them.  It is imperative to highlight that these assumptions are purely theoretical in nature. I neither assert that the practical development of a superhuman machine is feasible nor provide any prospective timeline for its realization.

Using this framework, I establish an impossibility theorem, which indicates that when these four assumptions are considered together, they lead to a contradiction. Put differently, under these assumptions, it becomes impossible for $M$ to surpass $H$ in two-person general-sum games.  The significance of this theoretical finding is twofold: (i) the framework encapsulates both non-zero-sum and zero-sum games, and (ii) the impossibility theorem is ``tight'' in its formulation, wherein the relaxation of even one assumption reinstates the consistency of the entire set. Through the identification and analysis of these assumptions and their inherent inconsistencies, this research offers a deeper insight into the prospective development of superhuman AI.

From my research, two straightforward policy recommendations follow: first, we should control access to specific human data to maintain Strategic Unpredictability; and second, we should grant select policymakers and AI researchers access to superhuman machine research to ensure Access to Machine's Strategy holds. My analysis contributes to a better understanding of the theoretical context that can shape the development of superhuman AI. By examining the potential emergence of superhuman AI and proposing policy recommendations, this paper aims to foster a responsible and informed dialogue among political economic actors and AI researchers and practitioners.

\subsection{Literature review}
\label{sec:literature}

The emergence of superhuman AI poses unprecedented challenges and risks for humanity. Many scholars have warned about the possible dangers of creating artificial agents that surpass human intelligence and capabilities. Some of these dangers include job automation and rising inequality \citep{ford2015,brynjolfsson2014}, security and privacy breaches, AI malware \citep{brundage2018}, autonomous weapons \citep{scharre2019}, deepfakes, fake news, and political instability \citep{chesney2019}. 

This paper builds on the existing literature that explores the emergence and potential threats posed by superhuman AI. This literature is vast and diverse, but some notable contributions include \citep{yudkowsky2008,bostrom2014,russell2019} and \citep{cohen2022} (henceforth CHO). CHO argue that advanced artificial agents are \textit{likely} to manipulate or interfere with their reward function, which could lead to disastrous outcomes due to conflicts of interest over resources between humans and advanced machines. My Superhuman Machine assumption is related to CHO's Assumption 6: ``A sufficiently advanced agent is likely to be able to beat a suboptimal agent in a game, if winning is possible.'' I extend this concept beyond games where winning is easily defined, such as zero-sum games, to include non-zero-sum games, where cooperation is not only possible but also common, and there is no clear-cut definition of victory or defeat. Furthermore, my concept of Strategic Unpredictability resembles CHO's ``self-sufficient model'' where human actions are simulated in the model. However, my approach differs from CHO's in two key ways. Firstly, I provide formal game theoretical definitions for my assumptions. Secondly, I interpret my assumptions from a political economy perspective.

The main theorem in this paper is a no-go theorem, which is a theorem that shows that specific physical or mathematical phenomena are precluded from occurring under particular conditions or assumptions. These theorems are typically used in theoretical physics to constrain the possible outcomes of a physical system. One no-go theorem pertinent to (quantum) computing is the no-cloning theorem \citep{wootters1982}, which roughly implies that quantum computers cannot simply make copies of any qubits, as is possible in classical computing.\footnote{It should be noted that versions of no-cloning theorems also exist within classical mechanics \citep{daffertshofer2002}.} \textit{If} the human brain operates according to the principles of quantum mechanics, then the no-cloning theorem would make it more difficult, or even impossible, to violate the Strategic Unpredictability assumption.

As mentioned above, superhuman $M$ is a theoretical construct that does not imply any practical possibility or timeline for its creation. However, the field of AI has witnessed remarkable achievements towards building superhuman machines in various domains over the last three decades. For instance, IBM's Deep Blue was the first chess engine to defeat a World Chess Champion, Garry Kasparov, in a match in 1997 \citep{campbell2002}. In 1992, \citet{tesauro1992} developed TD-Gammon at IBM, which was the first self-learning computer program that surpassed average human-level performance in a major board game. However, it was still inferior to the best human players at that time. Later, programs such as Backgammon Snowie and GNU Backgammon improved upon TD-Gammon's algorithm and achieved superhuman play in backgammon. More recently, DeepMind's AlphaGo was the first program to beat a top professional player in Go \citep{silver2016}. \citet{silver2018} introduced AlphaZero, which achieved superhuman play in not only one game, but in three different games: chess, shogi, and Go. In poker, \citet{brown2019} introduced the first program that achieves superhuman performance in six-player no-limit Texas hold'em. 

It should be noted that while some combinatorial games such as Nim have analytical solutions that do not require significant computing power to find optimal solutions, others such as Catch-Up do not have analytical solutions, and empirical evidence suggests that its optimal outcome may be a draw whenever possible \citep{isaksen2015}. Additionally, some games like Hex have not been solved analytically, but it can be shown that the first player has a winning strategy by a strategy-stealing argument. Although \citet{schaeffer2007} showed that checkers is a draw with optimal play from both players, it is unlikely that other major games such as chess and Go can be solved in the same way anytime soon due to their complexity.

\section{The setup and results}
\label{sec:setup}
\subsection{Non-zero-sum games}

\begin{table}[h!]
\[
\arraycolsep=1.1pt\def\arraystretch{1.3}
\begin{array}{ r|c|c|c|}
\multicolumn{1}{r}{}
&  \multicolumn{1}{c}{\text{Name}}
& \multicolumn{1}{c}{\text{Notation}}
& \multicolumn{1}{c}{\text{Element}}\\
\cline{2-4}
&  \text{Players}  & N=\{1,2\}  & i \\
&  \text{Nodes}  & X & x  \\
&  \text{Terminal nodes}  & Z& z \\
&  \text{Player function}  & I:X\to N &  \\
&  \text{Actions at node } x & A_i(x) & a_i(x)  \\
&  \text{Mixed strategy profiles}  & S & s  \\
&  ~\text{Probability mass on $a_i$ at $x$}~  & s_i(x)(a_i)  &   \\
& \text{Machine agent} & M& \\ 
& \text{Human agent} & H& \\ 
& \text{Superhuman Machine} & M^*& \\ 
& \text{Subgame at }x & G|x & \\
&  ~\text{Best-responses against}~s_j~ & BR_i(s_j) & s^*_i \\
& \text{Expected payoff of}~i~ & u_i:S\rightarrow \mathbb{R}& \\ 
&  \text{Extensive form game} & ~G=(N, X, I, u, S) ~&  \\
&  k\text{-repeated contest} & G^k_{1,2} &  \\
& \text{Sample average} & \mu(G) &  \\
\cline{2-4}
\end{array}
\]
\caption{A summary of the notation}
\label{table:terminology}
\end{table}

Table~\ref{table:terminology} introduces the notation I use in this paper.
Let $G=(N, X, I, u, S)$ be an extensive form game with perfect information and perfect recall, where $N=\{1,2\}$ is the set of players, $X$ a finite game tree with a node $x\in X$,  $x_0$ the root of the game tree, $z\in Z$ a terminal node, $I:X\setminus Z \rightarrow N$ the player function that assigns an active player to each non-terminal node, and $u$ the profile of payoff functions. For every player $i\in \{1,2\}$, $A_i(x)$ denotes the finite set of pure actions of player $i$ at node $x$ and $A_i=\bigcup_{x | I(x)=i}A_i(x)$ denotes the finite set of all pure actions of player $i$.

A pure strategy $s'_i$ of player $i$ is a function  $s'_i:X_i\rightarrow A_i$ such that $x\in X_i$, $s'_i(x)\in A_i(x)$, where $X_i$ is the set of nodes in $X$ where player $i$ acts. Let $S'_i=\bigtimes_{x| I(x)=i} A_i(x)$ denote the set of all pure strategies of $i$, and $s'\in S'=\bigtimes_{i\in N} S'_i$ a pure strategy profile. A mixed strategy $s_i$ of player $i$ is a probability distribution over $S'_i$, and $S_i= \Delta (S'_i)$ is the set of all mixed strategies of player $i$. Let $s\in S$ denote a mixed strategy profile and $s_i(x)(a_i)$ denote the probability with which player $i$ chooses action $a_i$ at node $x$. Player $i$'s (von Neumann-Morgenstern) expected payoff function is $u_i:S\rightarrow \mathbb{R}$. Let $s^*_i\in BR_i(s_j)$ denote a \textit{best-response} of player $i$ to player $j$'s strategy $s_j$, i.e., $s^*_i\arg\max_{s'_i\in S_i} u_i(s'_i,s_j)$.

$G$ is two-player game played between a representative human agent, denoted by $H$, and a machine agent, denoted by $M$. I use $s_{-i}$ to denote the strategy of player $j\neq i$. For any non-terminal node $x\in X$, I use $G|x$ to denote the subgame of $G$ whose game tree starts at node $x$ and contains all successor nodes in $X$. Similarly, I use $(s|x)$ to denote the strategy profile $s$ restricted to the subgame $G|x$.

\subsection{Concepts}

In game theory, a Nash equilibrium is a strategy profile in which no player can unilaterally improve their payoff holding the strategies of the others fixed. Formally, its definition is given as follows.

\begin{definition}[\citeauthor{nash1951}, 1951]
\label{def:Nash}
A strategy profile $s\in S$ is called a \textit{Nash equilibrium} if for every player $i$ and for every $s'_{i}\in S_i$, $u_i(s)\geq u_i(s'_i,s_{-i})$.
\end{definition} 

A subgame perfect Nash equilibrium (SPNE) is a refinement of the Nash equilibrium concept, which requires that the Nash equilibrium holds not only in the game as a whole but also in every subgame.

\begin{definition}[\citeauthor{selten1965}, 1965]
	\label{def:subgame_perfection}
A strategy profile $s \in S $ is called a \textit{subgame perfect Nash equilibrium} (SPNE) if for every player $i$ and for every non-terminal $x\in X$ where $i=I(x)$, $u_i(s|x)\geq u_i(s'_i,s_{-i}|x)$ for every $s'_{i}|x\in S_i|x$.
\end{definition}

To define the nature of competition between $H$ and $M$, I introduce the following definition.

\begin{definition}[Repeated contest]
	\label{def:repeated_contest}
	Let $G_1$ denote a game of $G$ in which player 1 is $H$ and player 2 is $M$, and $G_2$ denote a game of $G$ in which player 1 is $M$ and player 2 is $H$. Let $G^k_{1,2}$, $k\in \{1,2,...\}$, denote the \textit{repeated contest} game in which each stage game consists of two games, $G_1$ and $G_2$, and each stage game is repeated $k$ times.
\end{definition}

In simple words, the repeated contest between $H$ and $M$ is defined as the repeated game in which each stage game consists of two games in each of which the roles of the players are swapped. This is done to account for the possibility that  game $G$ may be biased towards one player. For example, in the world chess championship, the players play an equal number of games with white pieces to account for any potential first-mover advantage. I formalize the concept of outperformance as follows.

\begin{definition}[Outperformance]
\label{def:outperformance}
Let $G$ be a two-person perfect information game, $G^k_{1,2}$ the repeated contest, and $s_j$ be the player $j$'s strategy in $G^k_{1,2}$. Player $i\in \{H,M\}$ is said to \textit{outperform} player $j\neq i$ if there exists $\bar{s}_i\in S_i$ such that for any $k\in \{1,2,...\}$, $u_i(\bar{s}_i,s_j)>u_j(\bar{s}_i,s_j)$. 
\end{definition}

In plain words, player $i$ outperforms player $j$ in game $G$ if, no matter how many times the contest is repeated, player $i$'s expected payoff is strictly greater than player $j$'s.\footnote{Definition of outperformance can be extended to imperfect information games by restricting $k$ above a certain threshold, which depends on the game being played.} However, the number of repetitions needed to determine the ``better'' player in practice may depend on the specific characteristics of game $G$. To give an example, in a world chess championship match between two players, 20 repetitions may suffice to accurately determine the better player. On the other hand, in a backgammon championship, the contest must be repeated more times to accurately determine the better player.

\subsection{Assumptions}
\label{sec:assumptions}

\subsubsection{Superhuman machine}

I define a `superhuman' artificial intelligence, denoted by $M^*$, as an artificial general intelligence that is equipped with finite but significant computing power, and is able to take any game $G$ as an input and output a \textit{solution}---i.e., a mixed strategy profile--- based on its source code and computational power. While $M^*$ may not always be able to find an `optimal' solution for very large games, it can analyze the game tree and come up with a solution. Updating its solution as the game proceeds is also possible, similar to chess engines.

Determining whether a machine is `human-like' or `superhuman' is a subjective matter that involves human judgments, such as the well-known Turing test \citep{turing1950}. To define a superhuman machine, I first introduce a useful concept, namely the sample average of a two-player game played by a population of human agents.

\begin{definition}[Sample average]
\label{def:average}
Consider a population of human agents playing a two-player game $G$, and let $\{(u^1_1,u^1_2),(u^2_1,u^2_2),...,(u^n_1,u^n_n)\}$ be the dataset of payoffs, where $(u^j_1,u^j_2)$ is the payoff received by player 1 and player 2 from the $j$th game of $G$. It is possible that each game is played by different players. Then, the sample average is defined as follows:
\[
\mu(G) = \frac{1}{2n}\sum^{n}_{j=1} (u^j_1+u^j_2).
\]
\end{definition}

The sample average $\mu(G)$ of a game $G$ is determined by the empirical average  payoff received by a group of human players who participate in playing the game. The sample average can be obtained from a tournament that is designed and agreed upon by a group of experts in the game of $G$. These experts could either be experienced players or judges (e.g., a boxing judge) who have knowledge of the game but do not necessarily play it. In this paper, I assume that the sample average for a game $G$ is based on established empirical research, if any, on $G$.

I next introduce the definition of a superhuman machine.

\begin{definition}[Superhuman]
	\label{def:superhuman}
 A machine $M$ is called \textit{superhuman} if 
\begin{enumerate}
    \item  there exists $G'$ such that $M$ outperforms $H$ in $G'$,
    \item for every $G$, $M$ is not outperformed by $H$ in $G$, and
    \item for every $G$, there exists a strategy of human agent $s_H$ such that given machine's strategy $s_M$,  $u_M(s_M,s_H)\geq \mu(G)$.
\end{enumerate}
\end{definition}

In simple terms, for an artificial intelligence to be classified as superhuman ($M^*$), it must outperform a human player ($H$) in some games and never be outperformed by $H$ in any game. Additionally, it should be possible for $M^*$ to receive a payoff no less than the sample average payoff from every game. While these first two conditions would be sufficient for defining superhuman machine in zero-sum games, the third condition is necessary in non-zero-sum games in which cooperation is not only possible but also common.  Therefore, to avoid aggressive machine strategies that aim to minimize the human's payoff while also minimizing their own payoff in non-zero-sum games, I introduce the third condition. This leads to the first assumption of my paper.

\begin{assumption}[Superhuman Machine]
	\label{asp:superhuman}
Superhuman Machine (\textbf{SHM}) holds if $M$ is superhuman.
\end{assumption}

\subsubsection{Access to Machine's Strategy}

This assumption requires that the human agent $H$ has the permission to access the strategy of the superhuman machine $M^*$. Formally, it is stated as follows.

\begin{assumption}[Access to Machine's Strategy]
\label{asp:access}
\textit{Access to Machine's Strategy} (\textbf{AMS}) is satisfied if for every game $G$ and at any node $x$ in game $G$, \text{H} takes $M^*$'s strategy $s_{M}\in S_{M}$ as given.
\end{assumption}

This assumption is crucial for analyzing the game-theoretic implications of superhuman machine intelligence, as it ensures that $H$ can take $M^*$'s strategy as given. This is similar to accessing to a chess engine during a game. However, the main theorem will demonstrate that this assumption alone does not prevent the emergence of superhuman machine intelligence in general-sum games; the other three assumptions are also required.

\subsubsection{Rationality}
\label{sec:rationality}
I use the standard rationality assumption, which refers to the idea that the human player ($H$) is acting in a way that is consistent with their own payoff function.

\begin{assumption}[Rationality]
	\label{asp:rationality}
	Player $H$ is rational if in every game $G$, for every strategy $s_M$ of $M^*$, $H$ chooses a strategy 
	\begin{equation}
	\label{eq:rationality}
	s^*_H\in\arg\max_{s'_H\in S_H} u_H(s'_H,s_M).
	\end{equation}
Rationality (\textbf{R}) is satisfied if player $H$ is rational.
\end{assumption}
In other words, $H$ chooses a strategy that maximizes their own expected utility given $M^*$'s strategy, which is feasible given the \textbf{AMS} assumption. 

\subsubsection{Strategic Unpredictability}

The Strategic Unpredictability assumption concerns the ability of agent $H$ to choose any strategy they would like to choose. In the context of this paper, I assume that $M^*$ cannot program $H$'s decisions in a way that would enable it to perfectly predict $H$'s actions either deterministically or non-deterministically.

\begin{assumption}[Strategic Unpredictability]
	\label{asp:freedom_mind}
Let $s'_H\in S_H$ be $M^*$'s prediction of $H$'s strategy in a game $G$. \textit{Strategic Unpredictability} (\textbf{SU}) holds if $H$ is free to choose a strategy $s_H\in S_H$ such that $s_H\neq s'_H$ regardless of $G$ in which $H$ has at least two pure strategies.
\end{assumption}

In other words, regardless of the code of $M^*$ about the strategy of $H$, $H$ can change their strategy in any way they wish. This assumption ensures that player $H$ has the freedom to act in an unpredictable manner and cannot be coerced to follow any specific course of action assumed by $M^*$.\footnote{This is a mild assumption since a human player, who has access to the strategy of $M^*$, can always change the strategy that $M^*$ assumes for them.}

\subsection{Centipede game}
\label{sec:centipede}

\begin{figure*}
	\centering
	\begin{tikzpicture}[font=\footnotesize,scale=1.2]
	\tikzstyle{solid node}=[circle,draw,inner sep=1.2,fill=black];
	\tikzstyle{hollow node}=[circle,draw,inner sep=1.2,fill=black];
	\node(0)[solid node]{}
	child[grow=down]{node{}edge from parent node[left]{$\bar{S}$}}
	child[grow=right]{node(1)[solid node]{}
		child[grow=down]{node{}edge from parent node[left]{$\bar{S}$}}
		child[grow=right]{node(2)[solid node]{}
			child[grow=down]{node{}edge from parent node[left]{$\bar{S}$}}
			child[grow=right]{node(3)[solid node]{}
				child[grow=down]{node{}edge from parent node[left]{$\bar{S}$}}
				child[grow=right]{node(4){}
					edge from parent node[above]{$C$}
				}
				edge from parent node[above]{$C$}
			}
			edge from parent node[above]{$\cdots$}
		}
		edge from parent node[above]{$C$}
	};
	\foreach \x in {0,2}
	\node[above]at(\x){1};
	\foreach \x in {1,3}
	\node[above]at(\x){2};
	\node[below]at(0-1){$\begin{pmatrix} 2 \\ 1 \end{pmatrix}$};
	\node[below]at(1-1){$\begin{pmatrix} 1 \\ 4 \end{pmatrix}$};
	\node[below]at(2-1){$\begin{pmatrix} 2k_1 \\ 2k_1-1 \end{pmatrix}$};
	\node[below]at(3-1){$\begin{pmatrix} 2k_2-1 \\ 2k_2+2 \end{pmatrix}$};
	\node[right]at(4){$\begin{pmatrix} 2k_2+2 \\ 2k_2+1 \end{pmatrix}$};
	\end{tikzpicture}
	\caption{Payoff function of a linearly increasing-sum centipede game.}
	\label{fig:centipede}
\end{figure*}
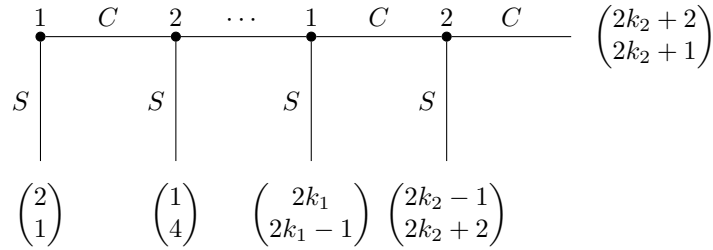

I next define a well-studied experimental game that will be useful to prove the main theorem. The centipede game of \citet{rosenthal1974} is a two-person perfect information game in which each player has two actions, continue (C) or stop ($\bar{S}$), at each decision node. There are several variations of this game, but some of the main characteristics of a standard centipede game include (i) the size of the ``pie'' increases as the game proceeds, (ii) if player $i$ chooses C at a node, then the payoff of player $j\neq i$ increases, and (iii) the unique subgame-perfect equilibrium is to choose $\bar{S}$ at every node. For example, suppose that there are $m\geq 2$ (even) decision nodes and let $k_i\in \{1,2,...,\frac{m}{2}\}$ be the node such that player $i$ is active. Figure~\ref{fig:centipede} illustrates the payoff structure of a linearly increasing-sum centipede game due to \citet{aumann1998}.

There have been numerous experimental studies on the centipede game and its variations since the work of \citet{mckelvey1992}. These studies include, among others, \citet*{fey1996}, \citet{nagel1998}, \citet{rubinstein2007}, \citet*{levitt2011}, and \citet*{krockow2016}, which is a meta-analysis of nearly all published centipede experiments. The most widely replicated finding is that in increasing-sum centipede games, human subjects tend to overwhelmingly choose to continue in their first opportunity and do not choose to stop, whereas in constant-sum centipede games, they mostly choose to stop in the first opportunity. Furthermore, as the length of the game increases, subjects tend to choose to stop later in increasing-sum centipede games (see, e.g. McKelvey and Palfrey, 1992).

The centipede game mean stopping node, defined by \citet{krockow2016}, is used to measure the average level of cooperation in centipede experiments. To account for the varying game lengths in experimental games, the mean stopping node is standardized by dividing it by the length of the game. The empirical evidence presented in \citeauthor{krockow2016}'s meta-analysis indicates that in linearly increasing-sum centipede games, the minimum standardized mean stopping node is 0.4 \citep[p. 246]{krockow2016}. In the following lemma, I show the sample average in centipede games.

\begin{lemma}[Sample average lower bound]
\label{lemma:centipede_average}
In linearly increasing-sum centipede games, the sample average satisfies the following condition: $\mu(G)>0.8m-0.5$. 
\end{lemma}

\begin{proof}
According to the meta study conducted by \citet{krockow2016}, the minimum standardized mean stopping node in linearly increasing-sum centipede games is $0.4$. Let $m$ be the length of the centipede game shown in Figure~\ref{fig:centipede}. At the minimum standardized mean stopping node, player 1 and player 2's payoffs are $0.8m$ and $0.8m-1$, respectively, resulting in an average payoff of $0.8m-0.5$. As $\mu(G)$ represents the sample average payoff of all players in the population, and $0.8m-0.5$ is the average payoff at the \textit{minimum} standardized mean stopping node in centipede games, it implies that the sample average payoff of all players in the population must be greater than the minimum average payoff, that is, $\mu(G)>0.8m-0.5$.
\end{proof}

\subsection{Results}

First, it is helpful to explicitly state what I mean by consistency.

\begin{definition}[Consistency]
\label{def:consistency}
A set assumptions are called \textit{consistent} if they do not lead to any logical contradiction. They are called \textit{inconsistent} if they are not consistent.
\end{definition}

The following no-go theorem shows that the existence of a superhuman $M$ is impossible if the three main assumptions hold.

\begin{theorem}[Impossibility of $M^*$]
	\label{thm:impossibility}
The assumptions \textbf{SU}, \textbf{AMS}, \textbf{R} and \textbf{SHM} are inconsistent. 
\end{theorem}

\begin{proof}
Assuming that \textbf{SU}, \textbf{AMS}, and \textbf{R} hold and that $M^*$ exists, I will prove by contradiction that $H$ outperforms $M^*$ in an increasing-sum centipede game $G$ whose length $m\geq 6$. 

To begin, let $s\in S$ be $M^*$'s solution in game $G$, defined by the payoff function in Figure~\ref{fig:centipede}. Suppose that $s_i|g\in BR_j(s_j|g)$ for every $i$ and every subgame $g$ of $G$, meaning that $M^*$ assigns best responses to each player at every decision node. Then, $s$ must be the unique subgame perfect Nash equilibrium in $G$, or else it would assign a non-best response to at least one player at one of the nodes. This relies on a well-known backward induction argument: in the last node $M^*$ must assign $\bar{S}$ to the active player, who might be $M^*$ or $H$, and given that $M^*$ must assign $\bar{S}$ in the last node, $M^*$ must assign $\bar{S}$ in the second-to-last node, and so on. Since $M^*$ is superhuman by Definition~\ref{def:superhuman}, this implies a contradiction to the \textbf{SHM} assumption, because choosing $\bar{S}$ in the first two nodes implies that $u_{M^*}(s)<\mu(G)$ by Lemma~\ref{lemma:centipede_average}, that is, $M^*$ receives strictly less than the sample average.

Suppose $s_{M}(x_0)(\bar{S})> 0.75$, meaning that $M^*$ assigns a probability of more than $0.75$ to choosing $\bar{S}$ at the root of the game. In this case, the maximum payoff $M^*$ can receive is less than $2\times 0.75+(m+2)\times 0.25$, where $m$ is the number of decision nodes in $G$. It implies that for any $m$, $2+0.25m < 0.8m-0.5$ if and only if $m>4.54545$. This means that for every $m>4$ and every $s'_H$, $u_M(s_M,s'_H)<\mu(G)$. Put differently, for a large enough $m$, it is impossible for $M^*$ to receive the sample average payoff in $G$.
As a result, it must be that $s_{M}(x_0)(C)\geq 0.25$, implying that $M^*$ chooses C at the root with a probability greater than $0.25$.

By the \textbf{AMS} assumption, $H$ takes the strategy $s_{M}$ of $M^*$ as given. Then, \textbf{R} implies that $H$ chooses a strategy in $\arg\max_{s'_i\in S_i} u_i(s'_i,s_{M})$, i.e., $H$ best-responds to the strategy of $M^*$. Define $\bar{s}_H\in \arg\max_{s'_i\in S_i} u_i(s'_i,s_{M})$ such that $\bar{s}_H$ is a pure strategy. Furthermore, the \textbf{SU} assumption implies that $H$'s strategy cannot be predicted by $M^*$, so $s_{M}\notin BR_{M}(\bar{s}_H)$. In other words, $M^*$'s strategy cannot be a best-response to $H$'s strategy because $H$ is already best-responding to $M^*$. If both players are best-responding to each other, then the only possible outcome is to choose $\bar{S}$ at the first node, which leads to a contradiction as shown above.

The payoff function of $G$ ensures that the player who best-responds with a pure strategy receives a greater payoff than the other player because $2k_1>2k_1-1$ and $2k_2+2> 2k_2-1$---unless player 1 chooses $\bar{S}$ with a high enough probability ($> 0.75$) at the root of the game, which is ruled out by the above argument. Therefore, $H$ outperforms $M^*$ in the repeated contest $G^k_{1,2}$ for any $k>0$ because in both $G_1$ (the game in which $H$ is player 1) and $G_2$ (the game in which $H$ is player 2), $u_H(\bar{s}_H, s_{M})>u_{M}(\bar{s}_H, s_{M})$. This implies that $H$'s payoff must be strictly greater than $M^*$'s payoff in the repeated contest.

As desired, $H$ outperforms $M^*$ in the repeated contest, which contradicts to the supposition that $M^*$ is superhuman.
\end{proof}

The proof strategy can be explained in simpler terms in seven main steps. 
\begin{enumerate}
    \item To reach a contradiction, suppose that \textbf{SU}, \textbf{AMS}, \textbf{R}, and \textbf{SHM} all hold.
    \item If $M^*$'s solution $s$ is an SPNE in the centipede game, then \textbf{SHM} must be violated due to Lemma~\ref{lemma:centipede_average}.
    \item Now suppose that $M^*$ stops at the first node with a high probability (but strictly below 1). But then it would be impossible for $M^*$ to receive the average sample payoff in the centipede game.
    \item Therefore, $M^*$ must choose C at the root with a high enough probability to receive the average sample payoff.
    \item Note that $H$ takes the strategy $s_M$ of $M^*$ as given by \textbf{AMS}, $H$ chooses a pure best-response to the strategy of $M^*$ by \textbf{R}, and $M^*$ cannot predict $H$'s strategy by \textbf{SU}. 
    \item These assumptions imply that $H$ outperforms $M^*$ in the repeated contest $G^k_{1,2}$ for any $k$ because  whether $H$ is the first player or the second player, $H$ receives a strictly greater payoff than $M^*$.
    \item Therefore, a contradiction is obtained. \textbf{SU}, \textbf{AMS}, and \textbf{R} imply that \textbf{SHM} does not hold.
\end{enumerate}

I next explore the ``tightness'' of Theorem~\ref{thm:impossibility} as mentioned earlier.

\begin{proposition}
\label{prop:tightness}
Theorem~\ref{thm:impossibility} is tight: Any three of the four assumptions, \textbf{SU}, \textbf{AMS}, \textbf{R}, and \textbf{SHM}, are consistent.
\end{proposition}

\begin{proof}
To prove this proposition, I drop each of the four assumptions \textbf{SU}, \textbf{AMS}, \textbf{R}, and \textbf{SHM} one by one and show that the remaining three assumptions do not lead to any contradictions.

\vspace{0.2cm}
\noindent \textit{Superhuman Machine}: I begin by assuming that \textbf{AMS}, \textbf{SU}, and \textbf{R} hold, but \textbf{SHM} does not. This is the easiest case, as there is no restriction on the behavior of the machine under these assumptions. Thus, these three assumptions are consistent.

\vspace{0.2cm}
\noindent \textit{Access to Machine's Strategy}: 
Assuming that \textbf{SU} and \textbf{R} hold but \textbf{AMS} does not hold, $H$ would best-respond to some belief about $M^*$'s strategy. However, there would be no guarantee that $H$'s belief is correct, which means $H$ would not necessarily be able to outperform $M^*$. This implies that $M^*$ \textit{may} be superhuman. Therefore, \textbf{SU}, \textbf{R}, and \textbf{SHM} are consistent.

\vspace{0.2cm}
\noindent \textit{Rationality}: 
Assume that \textbf{SU} and \textbf{AMS} hold, but \textbf{R} does not. Then, this assumption would \textit{not} contradict the assumption that $M$ is superhuman. This is because if $H$ fails to act rationally, then they may select a strategy that leads to being outperformed by $M^*$, which is consistent with the \textbf{SHM} assumption. As a result, \textbf{SU}, \textbf{AMS}, and \textbf{SHM} are consistent.

\vspace{0.2cm}
\noindent \textit{Strategic Unpredictability}: 
Assuming that \textbf{AMS} and \textbf{R} hold, but \textbf{SU} does not, $M^*$ might be able to program $H$'s brain and predict precisely what $H$ will choose and can best respond. This implies that one of the players could outguess the other player, depending on perhaps the computational power of $M^*$. As a result, one cannot rule out the scenario that $M^*$ outperforms $H$ in every game, in which case the theorem would not hold. This implies that \textbf{AMS}, \textbf{R}, and \textbf{SHM} are consistent.
\end{proof}

\section{Discussion and conclusions}
\label{sec:discussion}

This paper examines the emergence of superhuman artificial intelligence (AI) through a game theoretical perspective, considering the factors that could impact its development. Using a non-zero-sum framework to model strategic interactions between a human agent and a potential superhuman machine agent, I show that under certain assumptions, it is not possible for superhuman AI to consistently outperform humans in non-zero-sum two-person games.

My analysis identifies four key assumptions underlying some of the arguments about the development of superhuman AI: Strategic Unpredictability, Access to Machine's Strategy, Rationality, and Superhuman Machine. I show that these assumptions are inconsistent when taken together and that this result is ``tight'' in the sense that relaxing any one of them results in a consistent set of assumptions. By identifying these assumptions and their inconsistencies, this paper contributes to a better understanding of the  context that can shape the theoretical development of superhuman AI.

It is worth noting that my analysis has some limitations. First, the proof of my main theorem depends on constructing a counterexample using the centipede game. However, this counterexample is not a `pathological' case, but rather an empirically validated example of a non-zero-sum game where humans can cooperate efficiently despite theoretical predictions. The proof could be generalized to other games where cooperation is crucial. Second, my analysis does not consider the possibility of multiple superhuman machines interacting with multiple human agents. This scenario may introduce new challenges for formalizing the cooperation and conflict between humans and the machines.

Despite its limitations, my analysis contributes to the ongoing theoretical debate about the emergence of superhuman AI by offering a formal game-theoretic framework for modeling potential strategic interactions between a human agent and a superhuman machine. By identifying the assumptions that underlie some of the arguments about the threats of superhuman AI, and showing their inconsistency when assumed together, this paper provides a new theoretical perspective on this issue.

\bibliographystyle{chicago}
\bibliography{references}

\end{document}